\documentclass[12pt]{article}
\usepackage{graphicx}
\usepackage{cite}
\usepackage{amsfonts}
\usepackage{amssymb}
\usepackage{latexsym}
\def\ee{\end{equation}}
\def\ba{\begin{eqnarray}}
\def\ea{\end{eqnarray}}

\def\bq{\begin{quote}}
\def\eq{\end{quote}}

 at 10truept

\newcommand{\beq}{\begin{equation}}
\newcommand{\eeq}{\end{equation}}
\newcommand{\beqa}{\begin{eqnarray}}
\newcommand{\eeqa}{\end{eqnarray}}
\newcommand{\bea}{\begin{eqnarray}}
\newcommand{\eea}{\end{eqnarray}}
\newcommand{\p}{\partial}
\newcommand{\al}{\alpha}
 
 \newcommand{\ep}{\epsilon}

\newcommand{\lmk}{\left(}
\newcommand{\rmk}{\right)}

\newcommand{\lle}{\left<}
\newcommand{\rgr}{\right>}
\newcommand{\lb}{\left|}
\newcommand{\rb}{\right|}

\newcommand{\overskrift}[1]{\vspace{6.0mm}\noindent\textbf{#1}\vspace{1.5mm}}

\def\lesssim{~\mbox{\raisebox{-.6ex}{$\stackrel{<}{\sim}$}}~}

\def\ltap{\ \raise.3ex\hbox{$<$\kern-.75em\lower1ex\hbox{$\sim$}}\ }
\def\gtap{\ \raise.3ex\hbox{$>$\kern-.75em\lower1ex\hbox{$\sim$}}\ }
\def\gl{\ \raise.5ex\hbox{$>$}\kern-.8em\lower.5ex\hbox{$<$}\ }
\def\roughly#1{\raise.3ex\hbox{$#1$\kern-.75em\lower1ex\hbox{$\sim$}}}

\begin{document}

\thispagestyle{empty}
%\begin{titlepage}
%%%%%%%%%%%%%%%%%%%%%%%%%%%%%%%%%%%%%%%%%%%%%%%%%%%%%%%%%%%%%%%%%%%%%%%%
%%%%
%\noindent
\begin{flushright}
{\tt astro-ph/0604488}\\
%January 9, 2006
\end{flushright}

\vskip2cm
\begin{center}
{\Large{\bf On the One Loop Corrections to Inflation}}\\ {\Large{\bf and the CMB Anisotropies}} \\
\vskip2cm {\large Martin S. Sloth\footnote{\tt
sloth@physics.ucdavis.edu}}\\

\vspace{.5cm}

\vskip 0.1in

{\em Department of Physics, University of California}\\
{\em Davis, CA 95616, USA}\\

\vskip 0.1in
\vskip 0.1in
\vskip .25in
{\bf Abstract}
\end{center}
We investigate the one loop effective potential of
inflation in a standard model of chaotic inflation. The leading
one loop corrections to the effective inflaton potential are
evaluated in the quasi de Sitter background, and we estimate the
one loop correction to the two-point function of the inflaton
perturbations in the Hartree approximation. In this approximation,
the one loop corrections depends on the total number of e-foldings
of inflation and the maximal effect is estimated to be a correction to the
power spectrum of a few percent. However, such a correction may be
difficult to disentangle from the background in the simplest scenario.

%\end{abstract}

\vfill \setcounter{page}{0} \setcounter{footnote}{0}
\newpage
%\end{titlepage}

%%%%%%%%%%%%%%%%%%%%%%%%%%%%%%%%%%%%%%%%%%%%%%%%%%%%%%
%%%%%%%%%%%%%%%%%%%%%%%%%%%%%%%%%%%%%%%%%%%%%%%%%%%%%%
\setcounter{equation}{0} \setcounter{footnote}{0}
%%%%%%%%%%%%%%%%%%%%%%%%%%%%%%%%%%%%%%%%%%%%%%%%%%%%%%
%%%%%%%%%%%%%%%%%%%%%%%%%%%%%%%%%%%%%%%%%%%%%%%%%%%%%%

\section{Introduction}

The measurements of the Cosmic Microwave Background (CMB)
anisotropies proceed with increasing accuracy, which together with
Super Novae (SN) and large scale structure (LSS) data yields an
ever more detailed picture of the evolution of the universe.
Inflation has earned its place in this picture as a successful
paradigm, which, in addition to solving the theoretical problems
of the standard Big-Bang theory, explains a number of
observational data \cite{inflation}. However, we know little about
the theory of inflation itself, essentially due to our ignorance
of physics at the very short distance scales. In fact, even the
energy scale of inflation is at present unknown within $10$ orders
of magnitude. Future experiments might change that situation
dramatically. Especially a detection of primordial gravitational
waves could pin down the energy scale at which inflation takes
place, and as we measure the spectrum of CMB perturbations to
higher and higher accuracy we will soon be able to discriminate
between different models of inflation and rule out some of the
popular ones \cite{Spergel:2006hy}. This is especially exciting
because it is one of our few windows to fundamental physics well
above the TeV scale.

Naturally there is a large interest in the theoretical limitations
to what we can learn about the physics of inflation and thus
physics at very short distance scales from CMB measurements. In
addition to attempts to reconstruct parts of the effective
potential of inflation from data \cite{pot}, it has also been
attempted to find possible signatures of the $UV$ scale where the
effective theory of inflation breaks down, using modified
dispersion relations \cite{transpl}, an effective minimum length
\cite{minlength}, non-commutativity \cite{noncomm}, a new-physics
hyper-surface \cite{newphys}, or effective field theory
\cite{Kaloper:2002uj,eff,collins,Weinberg:2005vy}, combined with
constraints from more theoretical considerations \cite{holo}.
While these considerations are important, it is not yet clear
whether such effects are relevant in the real universe.

Here, we investigate the effect of one loop corrections to
inflation. If inflation has lasted very long, the Hubble rate at
the beginning of inflation $H_i$, is a new $UV$ scale in between
the Planck scale $M_p$ and the Hubble scale, $H$, when observable
modes exits the horizon. The physical modes corresponding to
length scales smaller than $1/H_i$, initially smaller than the
inflating patch, are pushed outside the horizon during inflation.
When integrated over in the loops, they might lead to a
significant enhancement of the one loop
effects\cite{vilenkin,woodard,Mukhanov:1996ak,Abramo:1997hu,Afshordi:2000nr,Finelli:2003bp,Losic:2005vg,Martineau:2005aa,Wu:2006xp}.
The physical $IR$ cutoff, given by the initial physical radius of
the inflating patch, is approximately equivalent to the apparent
particle horizon during inflation, and much larger than the Hubble
radius. This is the standard choice in the literature
\cite{vilenkin,woodard,Mukhanov:1996ak,Abramo:1997hu,Afshordi:2000nr,Finelli:2003bp,Losic:2005vg,Martineau:2005aa,Wu:2006xp}.
We expect that due to causality, to a local observer such $IR$
contributions can never uniquely be identified to be due
inhomogeneities, but will look very similar to an addition to the
background parameters of the effective theory and thus only
indirectly influence our measurements. On the other hand, the
effects are sensible to the $UV$ scale of the theory, which is the
Hubble rate at the beginning of inflation.

As an example, we will evaluate the one loop corrections to the
effective potential and the two-point function in a specific
$\lambda\phi^4$ type of chaotic inflation. We estimate that the
maximal effect in this particular model is a $1\%$ correction to
the power spectrum.

In the first part of the paper we will consider the effective one
loop potential and one loop corrections to the background
parameters of the theory. The effect appears to be maximally about
$1\%$. In the second part of the paper, we estimate the one loop
correction to the two-point function of the quantum fluctuations
in the Hartree approximation. We find a similar effect of order
$1\%$. This will lead to a small correction to the slow-roll
parameters and a possible change in the cosmological
scalar-to-tensor perturbation ratio, which is often also expressed
in the so called consistency relation.

Some of the earliest work on the one loop effects in inflation is
the work of Vilenkin \cite{vilenkin}, who studied the effective
potential of inflation in the approximation of an exact de Sitter
background. Some similar techniques to those applied in the second
half of the present paper, was applied in \cite{collins} in order
to understand the effect of a non-standard vacuum state on the
$UV$ cutoff on the one loop effects\footnote{An other approach to loop effects, the stochastic approach, has been applied by Linde \cite{Linde:1986fd} (see also \cite{Goncharov:1987ir}), in order to understand the global structure of space-time in eternal inflation.}. Finally, the effective
potential of inflation has been calculated in a slow-roll
approximation in \cite{Boyanovsky:2005sh,Boyanovsky:2005px}. They
found, as expected on dimensional grounds, that in an effective
field theory of inflation, one loop effects are suppressed by a
factor $H^2/M_p^2$ \cite{Kaloper:2002uj}, where $H$ is the
Hubble-rate when the observable modes exit and $M_p$ is the
reduced Planck scale. Our approach here is somewhat similar in
spirit. However, the effects we find are enhanced by a factor of
$H_i^4/H^4$, where $H_i$ is the Hubble-rate at the beginning of
inflation. This is because we go beyond the approximation of a
constant $H$. Our results are consistent with those of
\cite{Boyanovsky:2005px}, when we take $H_i\to H$.

Thus, it is not always that $H^2/M_p^2$ is the correct expansion
parameter for quantum effects in the effective field theory. In
long inflation there could be a significant enhancement, because
the loops are effectively suppressed by only a factor of
$H^4_i/(H^2M^2_p)$.

One might note, already in
\cite{Starobinsky:1992ts,Adams:2001vc,Kaloper:2003nv} it
was shown that even within the effective field theory approach, one
can have significant enhancements of quantum effects compared to the
dimensional $H^2/M_p^2$ estimate from potential bumps or other
non-adiabatic effects during inflation.

In the next subsection we will introduce the uniform curvature
gauge, which we will chose for our analysis in the next sections.
In section 2, we calculate the one loop corrections to the
equation of motion of the inflaton and the quantum corrections to
the slow-roll parameters. Our analysis is to first order in the
slow-roll and effective field theory expansion. In section 3, we
estimate the quantum corrections to the two point function of the
inflaton fluctuations in the Hartree approximation. In section 4,
we discuss our findings.

\subsection{Uniform curvature gauge}

We find that it is simplest to understand the one loop correction
to the inflaton potential in the uniform curvature gauge, because
here the residual degrees of the freedom of the perturbations can
be identified with the inflaton field fluctuations. This makes it
more transparent to understand the perturbations as quantum
fluctuations of the inflaton and to renormalize the divergences
with simple counter terms in the renormalized inflaton potential.

In order to calculate the renormalized two point function
$\lle\right. \delta\phi^2\left.\rgr_0$ in the effective theory, we
will thus write the perturbed metric to first order in the uniform
curvature gauge where the metric takes the form
\cite{Mukhanov:1990me,Hwang:1994rz}
 \beq
ds^2 =
-(1+2\varphi)dt^2+2aB_{,i}dtdx^i+a^2\delta_{ij}dx^{i}dx^{j}~.
 \eeq
In the absence of anisotropic stress $\dot B+2HB+2\varphi/a=0$, we
expect that there should be only one independent scalar degree of
freedom like in the longitudinal gauge if we have fixed the gauge
correctly. In fact one finds using the constraints from the
Einstein equations that the equation of motion for the inflaton
perturbations becomes
 \beq \label{eqmotx}
\ddot{\delta\phi}+3H\dot{\delta\phi}-\frac{1}{a^2}\nabla^2\delta\phi+\left(V_{\phi\phi}-6\ep
H^2 \right)\delta\phi=0~,
 \eeq
to first order in the slow-roll parameters. This shows that in
this gauge the quantum fluctuations can be identified with the
gauge invariant Sasaki-Mukhanov variable, $Q$, which satisfies the
same equation \cite{sasaki-mukhanov}. In appendix A, we have
generalized this to third order in perturbations and computed the
effective action of the perturbations to fourth order, since we
will need it to compute the self-interactions of the quantum
fluctuations of the inflaton.

Thus, in the zero curvature gauge we can describe the generation
of perturbations from quantum fluctuations of the inflaton in
Fourier space, by expanding the quantum field $\delta\phi(t,{\bf
x})$ in c-number mode functions with respect to the Bunch-Davis
vacuum
 \beq
\delta\phi(t,{\bf x}) = \int \frac{d^3{\bf
k}}{(2\pi)^{3/2}}\left[U_k(t)e^{i{\bf k\cdot x}}a_{\bf
k}+U^*_k(t)e^{-i{\bf k\cdot x}}a^{\dagger}_{\bf k}\right]~,
 \eeq
where the operator $a_{\bf k}$ annihilates the Bunch-Davis vacuum
$\lb 0\rgr_0$. The mode functions satisfy the Fourier transform of
eq.(\ref{eqmotx}), which in conformal coordinates yields
 \beq \label{eqmotk}
\left[\eta^2\frac{\p^2}{\p\eta^2}-2\eta\frac{\p}{\p\eta}+\eta^2
k^2+\frac{V_{\phi\phi}-6\ep H^2}{H^2}\right]U_k(\eta)=0~,
 \eeq
and the conformal time is defined as $a(\eta)d\eta=dt$. So, with
the usual normalization to the Minkowski vacuum in the infinite
past the solution for the modes becomes
 \beq
U_{k}(\eta)=\frac{\sqrt{\pi}}{2}H\eta^{3/2}H_{\nu}^{(2)}(k\eta)~,
 \eeq
where $H_{\nu}^{(2)}(k\eta)$ is the usual second Hankel function
and $\nu = 3/2+3\ep-\eta$. The spectral index of the perturbations
is defined as
 \beq
\mathcal{P}_{\delta\phi}(k,t) = \frac{k^3}{2\pi^2}\lle\right.
\left|\delta\phi_k(t)\right|^2\left.\rgr_0=\frac{k^3}{2\pi^2}\left|U_k(t)\right|^2~,
 \eeq
which on super-Hubble scales approximately gives
 \beq
\mathcal{P}_{\delta\phi}(k,t) \simeq
\frac{H^2}{4\pi^2}\left(\frac{k}{aH}\right)^{n-1}~,
 \eeq
with $n-1=3-2\nu$. It is important to note that this solution is
to first order in the slow-roll parameters and treating them as
constant.

\section{One loop effective field theory of inflation}

Inflation is most often formulated in terms of a minimally coupled
scalar inflaton quantum field, $\phi$, with a Lagrangian
 \beq
\mathcal{L} = -\frac{1}{2} (\p_{\mu}\phi)^2-V(\phi)~,
 \eeq
in a quasi de Sitter space with metric
 \beq
ds^2 = -dt^2 +a^2(t)\delta_{ij}dx^idx^j~,
 \eeq
where the scale factor is conveniently written on the form $a(t) =
\exp(Ht)$ and the Hubble rate $H$ is almost constant, a statement
which we can quantify in terms of the slow-roll parameter $\ep
=-\dot H/H^2 <<1$.

For definiteness we will consider a simple version of chaotic
inflation driven by an inflaton field with potential
 \beq
V(\phi) = \frac{1}{4}\lambda\phi^4~.
 \eeq
The resulting equation of motion is
 \beq\label{eqm}
\Box_g\phi+\lambda\phi^3 = 0~.
 \eeq
The dynamics is conveniently analyzed in an effective field theory
for the classical background field $\phi_c$, defined by
 \beq
\phi = \phi_c +\delta\phi~,
 \eeq
where $\delta\phi$ is a quantum field with a vanishing vacuum
expectation value enforced through the tadpole condition
 \beq
 \lle \delta\phi \rgr_0\equiv \lle 0\rb \delta\phi \lb 0\rgr = 0~,
 \eeq
such that
 \beq
\lle \phi \rgr_0\equiv \lle 0\rb \phi \lb 0\rgr = \phi_c~.
 \eeq
From Wick's theorem it follows that also
$\lle\right.\delta\phi^3\left.\rgr_0 = 0$. The effective equation
of motion for the classical field $\phi_c$ can then be obtained by
taking the expectation value of the equation of motion in eq.
(\ref{eqm}), which yields
 \beq
\lle\right.\Box_{g+\delta
g}\phi\left.\rgr_0+3\lambda\lle\right.\delta\phi^2\left.\rgr_0\phi_c+\lambda\phi_c^3
= 0~.
 \eeq
To calculate $\lle\right.\Box_g\phi\left.\rgr_0$, we need to take
into account also the metric perturbations in the uniform
curvature gauge as discussed in the introduction. The constraint
equations relating the scalar metric perturbations to the scalar
field fluctuations are in general complicated. Thus, we find that
the effective equation of motion to second order, is most easily
computed from the effective action of the perturbations expanded
to third order using the ADM formalism \cite{Arnowitt:1962hi}. In
the appendix we have in fact derived the effective action of the
perturbations to fourth order, which we will need later. The
expressions are given in eq.(\ref{S3}) and eq.(\ref{S4}). Using
the linear perturbation equation eq.(\ref{eqmotx}), we find from
eq.(\ref{S3}), on super-Hubble scales where we can neglect
gradient terms, to leading order
 \beq \label{effeqmot}
\ddot\phi_c+3H\dot\phi_c+6\lambda\lle\right.\delta\phi^2\left.\rgr_0\phi_c+\lambda\phi_c^3
= 0~.
 \eeq
The contributions from sub-Hubble scales, contributing to the
trace-anomaly, where computed in \cite{Boyanovsky:2005px}.
However, since they are not amplified by long inflation, those
contributions are always suppressed by a factor $H^2/M_p^2$ and we
will therefore ignore gradient terms and second order time
derivatives of the perturbations. Thus, to obtain an order of
magnitude estimate of the minimal one loop effects, it should be
sufficient to consider only the contributions in
eq.(\ref{effeqmot}). In section 3.2 we will see more formally how
the effective one loop equation of motion in eq.(\ref{effeqmot})
follows from the tadpole renormalization condition of the quantum
fluctuations.

From the effective equation of motion in eq.(\ref{effeqmot}), we
can see that the effective mass gets a one loop contribution
 \beq \label{meff}
\delta m^2_{eff} = 6\lambda\lle\right.\delta\phi^2\left.\rgr_0~.
 \eeq
The effective potential, which is consistent with the effective
energy density derived from the effective energy-momentum tensor
computed in \cite{Mukhanov:1996ak,Abramo:1997hu}, is
 \beq
V_{eff}(\phi_c)=
V(\phi_c)+2V'\lle\right.\delta\phi\varphi\left.\rgr_0+\frac{1}{2}V''\lle\right.\delta\phi^2\left.\rgr_0~.
 \eeq
The inflationary expansion is then given by the slow-roll of the
classical field in the effective potential, and in the slow-roll
approximation the effective Hubble rate is determined by the
effective Friedman equations
 \beq
3H^2 \simeq\frac{1}{M_p^2}V_{eff}(\phi_c)~,\qquad
3H\dot\phi_c\simeq -V'(\phi_c)-\delta m^2_{eff}\phi_c~,
 \eeq
as can be seen from the effective equation of motion
eq.(\ref{effeqmot}), with $\ddot\phi_c\simeq 0$. The effective
slow-roll parameters receives a one loop correction
$\delta\ep_{eff}$, $\delta\eta_{eff}$, when compared to the
tree-level slow-roll parameters $\ep$, $\eta$, such that
 \beq \label{effepeta}
\ep_{eff}=\ep+\delta\ep_{eff}~,\qquad \eta_{eff}=\eta
+\delta\eta_{eff}~.
 \eeq
With the following definition of the slow-roll parameter,
 \beq \label{effen}
\ep_{eff}\equiv \frac{1}{2M_p^2}\frac{\dot\phi_c^2}{H^2}~,
 \eeq
we can compute
 \beq \label{eeff}
\delta\ep_{eff}\simeq \frac{4}{3}\frac{\delta m_{eff}^2}{H^2}~,
 \eeq
and we assumed that the one loop contribution to the effective
potential is small compared to the tree-level value, i.e. $\delta
m_{eff}^2\phi_c^2<< V(\phi_c)$.

To make the approximations self-consistent, we should use the
effective slow-roll parameters in eq.(\ref{effepeta}) to calculate
the quantum fluctuations in eq.(\ref{eqmotx}). This is similar to
the \textit{cactus} or \textit{Hartree} approximation
\cite{vilenkin}, which corresponds to an infinite summation of
self-energy diagrams of the form shown if fig.(1). In section 3,
we shall calculate explicitly the one loop diagram in fig.(1),
which is first order in $\lambda$ and similar to computing the
loop correction to $\eta$ using Dyson's equation. To second order
in $\lambda$ there are also other one loop diagrams, not included
in the cactus approximation, that contributes, since the vacuum
expectation value of the background field is non-vanishing.

Assuming, according to the standard lore, that the initial vacuum
is the Euclidean vacuum, the spectral index $n_{eff}$ of the
primordial scalar curvature perturbations generated during
inflation follows
 \beq
n_{eff}-1 = 2\eta_{eff}-6\ep_{eff}~,
 \eeq
as we will discuss in more details in the next section where we
will also calculate the magnitude of the corrections. Especially
we need to calculate $\lle\right. \delta\phi^2\left.\rgr_0$ and
extract the dominant finite part.

\begin{figure}[!hbtp] \label{fig01}
\begin{center}
\includegraphics[width=6.3cm]{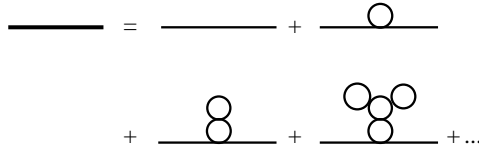}
\end{center}
\caption{The self-energy diagrams in the cactus approximation.}
\end{figure}

\subsection{Renormalized quantum fluctuations}

If one assumes that the inflationary state is preceded by a
radiation dominated one, then the expectation value $\lle\right.
\delta\phi^2\left.\rgr_0$ will receive two contributions. A
thermal contribution and a vacuum contribution. As inflation kicks
in, only the vacuum contribution will grow, and one can ignore the
thermal contribution. It is the modes corresponding to physical
wavelengths which are stretched to super-Hubble scales by the
inflationary expansion, that are responsible for the growth. As
modes are continuously pushed outside the horizon, they freeze and
give a little accumulating contribution to the expectation value.
Thus, the modes that where already outside the horizon at the
beginning of inflation, gives only a constant contribution, which
one can ignore a while after inflation has started.

Above, we calculated the one loop induced mass in terms of
$\lle\right. \delta\phi^2\left.\rgr_0$, which we can now compute
directly from the power spectrum
 \beq
\lle\right. \lb \delta\phi(t,{\bf x}) \rb^2\left.\rgr_0 =
\int\frac{d^3{\bf k}}{(2\pi)^{3/2}}\lle\right.
\left|\delta\phi_k(t)\right|^2\left.\rgr_0
=\int_{a_iH_i}^{a\Lambda}\frac{dk}{k}\mathcal{P}_{\delta\phi}(k,t)~.
 \eeq
The $IR$ cutoff on comoving momenta $a_iH_i$ is dynamically given
by the scale that exits the horizon at the beginning of inflation,
as explained above, and the $UV$ cutoff has been introduced by
hand to regulate a quadratic $UV$ divergence, which should be
subtracted by a proper renormalization. When the $UV$ divergences
has been subtracted, one finds that $\lle\right.
\delta\phi^2\left.\rgr_0$ is dominated by the $IR$ modes.

In order to actually evaluate the integral, we will split it in an
$IR$ part and a $UV$ part that we will evaluate separately
 \beq
\lle\right. \lb \delta\phi(t,{\bf x}) \rb^2\left.\rgr_0
=\lle\right. \lb \delta\phi(t,{\bf x}) \rb^2\left.\rgr_{IR}
+\lle\right. \lb \delta\phi(t,{\bf x}) \rb^2\left.\rgr_{UV}~,
 \eeq
with
 \beq
\lle\right. \lb \delta\phi(t,{\bf x}) \rb^2\left.\rgr_{IR}
=\int_{a_iH_i}^{a_0H_0}\frac{dk}{k}\mathcal{P}_{\delta\phi}(k,t)~,\qquad
\lle\right. \lb \delta\phi(t,{\bf x}) \rb^2\left.\rgr_{UV}
=\int_{a_0H_0}^{a\Lambda}\frac{dk}{k}\mathcal{P}_{\delta\phi}(k,t)~.
 \eeq
Here we let $a_0$, $H_0$ denote the values of $a$, $H$ at the time
when the relevant scales for the CMB crosses outside the horizon.

Let us first evaluate the more important $IR$ part. To properly
account for the running of the spectral index with the scale, we
will solve the mode function, $U_k$, in pure de Sitter, but using
the value of $H=H_k$ and $\dot\phi=\dot\phi_k$ when the given mode
crosses the horizon. If we specify to the specific $\lambda\phi^4$
model of chaotic inflation, this implies that we obtain the right
$k$ scaling from (See appendix B)
 \beq
U_k(\eta) = \lmk\frac{H_k}{H(\eta)} \rmk^{3/2}U^{ds}_k(\eta)=
\frac{\sqrt{\pi}}{2}H\eta^{3/2}\lmk\frac{H_k}{H(\eta)}
\rmk^{3/2}H_{3/2}^{(2)}(k\eta)~,
 \eeq
where the mode solution in pure de Sitter was denoted by
$U^{ds}_k(\eta)$.

If we let $N=\ln(a/a_i)$ denote the number of e-foldings of
expansions since the beginning of inflation, it is convenient to
recast the relevant integral into an integral over $N$. To do so,
we use the relations $d\ln k= dN$, $\ln(aH/k)=N$ and
 \beq
H_k^2 =\frac{\lambda}{12M_p^2}\left(\phi_i^2-64M_p^2N\right)^2~.
 \eeq
Now we can easily evaluate the dominant part of the $IR$
contribution to the relevant correlator, we obtain from
eq.(\ref{corr2}) in appendix B
 \beq
\lle\right. \lb \delta\phi(t,{\bf x}) \rb^2\left.\rgr_{IR} \simeq
\frac{H^2}{16\pi^2}\lmk\lambda\frac{1024}{3}\frac{M_p^2}{H^2}\rmk^{3/2}
N^4~.
 \eeq
If we had used same type of argument for a $m^2\phi^2$ theory, the
$IR$ part of the power spectrum would instead of $H_k^{3/2}$ scale
as $H_k^2$. If one applies the relation $\varphi = -
\sqrt{\ep/2}\delta\phi/M_p$, the result is consistent with
\cite{Mukhanov:1996ak,Abramo:1997hu,Afshordi:2000nr}.

The $UV$ part is much easier to evaluate. Inside the horizon, the
modes are not sensitive to the details of the expansion
 \bea
\lle\right. \lb \delta\phi(t,{\bf x}) \rb^2\left.\rgr_{UV}
&\simeq&
\frac{1}{8\pi}H^2\int_{1}^{\Lambda_H}\frac{dp_H}{p_H}p_H^3\left|H_{3/2}^{(2)}(p_H)\right|^2\nonumber\\
&=&\frac{1}{8\pi^2}H^2\left(\Lambda_H^2+\ln(\Lambda_H^2)\right)~.
 \eea
where $p_H=|k/(aH)|=|k\eta|$.

Finally we can add the $IR$ and the $UV$ contributions to obtain
the full correlator to leading order
 \bea
\lle\right. \lb \delta\phi(t,{\bf x}) \rb^2\left.\rgr_{0} &\simeq&
\frac{1}{8\pi^2}H^2\left(\Lambda_H^2+\ln(\Lambda_H^2) +
\Delta_N\right)~,
 \eea
where we for convenience have defined
 \beq \label{dn1}
\Delta_N
=\frac{1}{2}\lmk\lambda\frac{1024}{3}\frac{M_p^2}{H^2}\rmk^{3/2}
N^4~.
 \eeq
Another way of writing $\Delta_N$, which is illuminating, is in
terms of the Hubble rate at the beginning of inflation $H_i$,
which gives
 \beq \label{dn2}
\Delta_N =
\frac{\sqrt{3}}{64}\frac{1}{\sqrt{\lambda}}\frac{H_i^4}{M_pH^3}~.
 \eeq

The $UV$ divergent parts are canceled in the renormalized
potential introducing appropriate counter-terms. We shall not
elaborate on this aspect, but note that the physical $IR$
contributions survives as the dominant one loop correction to the
effective renormalized potential after subtracting the $UV$
divergences, as discussed in more details in
\cite{Boyanovsky:2005sh}.

\subsection{Quantum corrections to slow-roll parameters}

We can now evaluate the quantum corrections to the slow-roll
parameters and the spectral index of the CMB anisotropies, using
the relations derived in eq.(\ref{meff}) and eq.(\ref{eeff}). The
corrections to the slow-roll parameters can then be written in
terms of $\Delta_N$ in the following way
 \beq
\delta\ep_{eff}=\frac{\lambda}{\pi^2}\Delta_N~.
 \eeq
From eq.(\ref{dn2}) we obtain
 \beq
\delta\ep_{eff} =
\frac{\sqrt{3}}{64\pi^2}\sqrt{\lambda}\frac{H_i^4}{M_pH^3}~.
 \eeq
If we use the background values for the slow-roll parameters
 \beq
\ep = \sqrt{\frac{16\lambda}{3}}\frac{M_p}{H}~,
 \eeq
we can easily write the effective slow-roll parameter $\ep_{eff}$
in terms of a fractional correction to the background value
 \beq
\ep_{eff} = \ep\lmk 1+\frac{\delta\ep_{eff}}{\ep}\rmk = \ep\lmk 1
+\frac{3}{256\pi^2}\frac{H^2}{M_p^2}\frac{H_i^4}{H^4}\rmk~.
 \eeq
From an effective field theory point of view we generically expect
that corrections are of order $H^2/M_p^2$
\cite{Kaloper:2002uj,Boyanovsky:2005sh}. In fact, if we take
$H_i\to H$ that is exactly the type of correction we find.
However, due to the variation in the expansion rate over time, the
effect is amplified by the potentially large term $H_i^4/H^4$. It
is also interesting to note that
 \beq
\frac{H_i^4}{M_p^4} \simeq 10^5  \lambda^2 N^4~,
 \eeq
so the correction depends very sensitively on the total number of
e-foldings. This leads us to the possibility that the spectral
index carries a tiny imprint from the beginning of inflation
through the loop effects, such that we in principle indirectly can
observe the total number of e-foldings of inflation in this model.

In the model of chaotic inflation with a potential
$V(\phi)=1/4\lambda\phi^4$, in the regime when $V(\phi)>M_p^4$
quantum gravity effects are large and no classical description of
space is possible within the effective field theory framework.
When $10^{-4}M_p^4\lesssim V(\phi)\lesssim M_p^4$, the amplitude
of inflaton fluctuations is large compared to the mean and the
perturbations can not be treated perturbatively. This is the
self-reproduction regime. For our effective field theory
description of inflation to be consistent, we must therefore at
least require that the energy density of inflation, $\rho$, is
safely below the Planck scale,
 \beq
\rho^{1/4}\lesssim 0.1 M_p~.
 \eeq
This implies that the Hubble rate at the beginning of inflation
must satisfy $H_i^4\lesssim 10^{-9} M_p^4$. If we assume that the
Hubble rate, when the observable modes left the horizon, is given
by $H\simeq 10^{-5} M_p$, then the maximal correction to the first
slow-roll parameter, $\ep$, is
 \beq
\frac{|\delta\ep_{eff}|}{\ep}\lesssim \frac{30}{256\pi^2} \simeq
0.01~.
 \eeq
Thus, the maximal correction to the slow-roll parameter, $\ep$, is
less than about $1\%$ in this approximation.

\section{One loop contributions to the power spectrum}

In order to evaluate the one loop contributions self-consistently,
we need to compute the one loop contribution to the two-point
function of the quantum fluctuations. It is convenient to apply
the Schwinger-Keldysh formalism
\cite{Schwinger:1960qe,Keldysh:1964ud} for this purpose. Below, we
will first briefly review the Schwinger-Keldysh formalism and then
calculate the two-point function in the Hartree approximation.
This is similar to computing the one loop correction to the second
slow-roll $\eta_{eff}$ using Dyson's equation.

\subsection{Schwinger-Keldysh formalism}

This formalism has recently been reviewed and applied to the specific
problem of a self-interacting scalar field in de Sitter space,
in order to understand the $UV$ properties of inflaton
fluctuations originating in an alpha vacuum state \cite{collins}.
Here we will therefore only review it briefly in order to
introduce the proper notation, which will essentially follow
\cite{collins}.

The Schwinger-Keldysh formalism is a perturbative approach for
solving the evolution of a matrix element over a finite time
interval. This is useful in the inflationary coordinates of de
Sitter space, where there is no well defined asymptotic
\textit{in} and \textit{out} state in which to define an
\textit{S}-matrix, essentially due to the explicit time-dependence
of the metric. Instead of specifying an initial state in the
infinite past, one develops a given state forward in time from a
specified initial time $\eta_{infl}$, which we can think of as
being the beginning of inflation.

In the interaction picture, the evolution of operators is given by
the free Hamiltonian, $H_0$, while the part of the Hamiltonian
containing the interactions, $H_I$, is used to evolve the states
in the theory. If one specifies a density of state
$\rho(\eta_{infl})$ at a specific moment in time, then one can
define a unitary time evolution operator $U_I(\eta,\eta')$ that
evolves the state, and which is given by Dyson's equation
 \beq \label{Dyson}
U_I(\eta,\eta_{infl}) = T\left\{
e^{-i\int_{\eta_{infl}}^{\eta}d\eta'H_I(\eta')}\right\}~,
 \eeq
where $T$ is the time ordering of the product in the curly
brackets. Then one can write
 \beq
\rho(\eta) =
U_I(\eta,\eta_{infl})\rho(\eta_{infl})U_I^{-1}(\eta,\eta_{infl})~.
 \eeq
Absorbing a step function $\Theta(\eta-\eta_{infl})$ into the
interaction Hamiltonian, $H_I$, such that interaction only turn on
after $\eta_{infl}$, one can then write the evolution of the
expectation value of some operator, $\mathcal{O}$, as
 \beq \label{O1}
\lle\right.\mathcal{O}\left.\rgr(\eta)= \frac{\textrm{Tr}\left[
U_I(-\infty,0)U_I(0,\eta)\mathcal{O}U_I(\eta,-\infty)\rho(\eta_{infl})
\right]}{\textrm{Tr}\left[
U_I(-\infty,0)U_I(\eta,-\infty)\rho(\eta_{infl}) \right]}~.
 \eeq
This matrix element describes a system in the initial state
$\rho(\eta_{infl})$, evolved from conformal time $-\infty$ to $0$
with an operator inserted at $\eta$, and back again from $0$ to
$-\infty$. To evaluate this matrix element, one can formally
double the field content of the theory, with a set of "+" fields
on the increasing-time contour and a set of "-" fields on the
decreasing-time contour and then group the evolution operators
into a single time-ordered exponential. One can then write the
interacting part of the action appearing in Dyson's equation
together in a single time-contour, as
 \beq \label{SI}
S_I = -\int_{-\infty}^{0}d\eta\left[
H_I(\psi^+)-H_I(\psi^-)\right]~,
 \eeq
where contractions between different pairs of the two types of
fields now yields four kinds of propagators
 \bea
 \lle 0\rb
T\left[\psi^{\pm}(x)\psi^{\pm}(x')\right]\lb 0\rgr & = & -i G^{\pm\pm}(x,x')\nonumber\\
&=&-i\int\frac{d^3k}{(2\pi)^3}e^{i\vec{k}\cdot(\vec{x}-\vec{x}')}G_k^{\pm\pm}(\eta,\eta')~.
 \eea
The time-ordering of the contractions then yields
 \bea
G^{++}_k(\eta,\eta')&=&
G^{>}_k(\eta,\eta')\Theta(\eta-\eta')+G^{<}_k(\eta,\eta')\Theta(\eta'-\eta)\nonumber\\
G^{--}_k(\eta,\eta')&=&
G^{>}_k(\eta,\eta')\Theta(\eta'-\eta)+G^{<}_k(\eta,\eta')\Theta(\eta-\eta')\nonumber\\
G^{-+}_k(\eta,\eta')&=& G^{>}_k(\eta,\eta')\nonumber\\
G^{+-}_k(\eta,\eta')&=& G^{<}_k(\eta,\eta')~,
 \eea
where
 \bea
G^{>}_k(\eta,\eta')&=& iU_k(\eta)U^*_k(\eta')\nonumber\\
G^{<}_k(\eta,\eta')&=& iU^*_k(\eta)U_k(\eta')~.
 \eea
One can of course also define $G^{>}(x,x')$, $G^{<}(x,x')$ from
which $G^{>}_k(\eta,\eta')$, $G^{<}_k(\eta,\eta')$  can obtained
by a Fourier transform.

From Dyson's equation eq.(\ref{Dyson}) and eq.(\ref{SI}), one
finds that eq.(\ref{O1}) yields \cite{collins}
 \beq
\lle 0\rb \mathcal{O} \lb 0\rgr = \frac{\lle 0\rb
T\left\{\mathcal{O}~e^{-i\int_{-\infty}^{0}d\eta\left[
H_I(\phi_c,\psi^+)- H_I(\phi_c,\psi^-)\right]}\right\}\lb
0\rgr}{\lle 0\rb T\left\{e^{-i\int_{-\infty}^{0}d\eta\left[
H_I(\phi_c,\psi^+)- H_I(\phi_c,\psi^-)\right]}\right\}\lb 0\rgr}~,
 \eeq
if the initial state is the vacuum state $\lb 0\rgr$. Since we
absorbed the step function $\Theta(\eta-\eta_{infl})$ in $H_I$,
the time integral effectively have $\eta_{infl}$ as lower limit.

\subsection{Hartree approximation}

In appendix A, we have expanded the action for the inflaton field
fluctuations to fourth order. From the action given in
eq.(\ref{S3}) and eq.(\ref{S4}), it is simple to compute the
interaction Hamiltonian. Like in section 2, we shall constrain
ourself for simplicity, to consider only interactions that do not
contain space derivatives. These are the interactions that we
naively expect can give a large one loop contribution after a long
period of inflation. After using the results in eq.(\ref{S3}),
eq.(\ref{S4}), some partial integrations, and applying the linear
perturbation equation in eq.(\ref{eqmotx}), we obtain with
$\psi\equiv\delta\phi$, that the effective interaction Hamiltonian
in the present approximation can be given approximately as
 \bea
H_I(\phi_c,\psi^{\pm}) &\simeq& \int \frac{d^3y}{\eta^4
H^4}\left[\psi^{\pm}\lmk
\phi''_c+2\mathcal{H}\phi_c'+\lambda\phi_c^3\rmk\right.\nonumber\\
 & &\qquad+\left.2\lambda\phi_c{\psi^{\pm}}^3(y)
+\frac{15}{4}\lambda{\psi^{\pm}}^4(y)\right]~,
 \eea
in order to estimate the one loop correction to the two-point
function of the inflaton field fluctuations. Our equations are
consistent with section two. In fact, to first order in $\lambda$
one can verify that the tadpole renormalization condition
\cite{collins},
 \bea
0 &=& \lle \psi^{\pm}(x) \rgr_0 \nonumber\\
&=&-\int_{-\infty}^{\eta_0}d\eta\int\frac{d^3\vec{y}}{\eta^{4}H^4(\eta)}\left[\lmk
G^>(x,y)-G^<(x,y)\rmk\lmk
\phi''_c+2\mathcal{H}\phi_c'+\lambda\phi_c^3-i6\lambda\phi_c
G^>(y,y)\rmk\right]~.~,
 \eea
yields the effective one loop equation of motion in
eq.(\ref{effeqmot}).

The two-point function to one loop order can be organized in terms
of contributions to zero $T^{(0)}$, first $T^{(1)}$, second
$T^{(2)}$ and second order in $\lambda$,
 \beq
\left<\psi^+(\eta_0,\vec{x}_1),\psi^+(\eta_0,\vec{x}_2)\right> =
T^{(0)}(\eta_0,|\vec{x}_1-\vec{x}_2|)+T^{(1)}(\eta_0,|\vec{x}_1-\vec{x}_2|)+T^{(2)}(\eta_0,|\vec{x}_1-\vec{x}_2|)~,
 \eeq
where $T^{(0)}$ is the lowest order free tree-level contribution
to the two-point function. The first order contribution in
$\lambda$ receives a contribution
 \bea
T^{(1)}(x_1,x_2)
&=&i\int_{-\infty}^{\eta_0}d\eta\int\frac{d^3\vec{y}}{\eta^{4}H^4(\eta)}\lmk
G^>(x_1,y)G^>(x_2,y)-G^<(x_1,y)G^<(x_2,y)\rmk \nonumber\\
& &\times 45\lambda G^>(y,y)~.
 \eea
To second order in $\lambda$ there is two one loop contributions
 \beq
T^{(2)}(x_1,x_2)= T^{(2)}_1(x_1,x_2)+T^{(2)}_2(x_1,x_2)~,
 \eeq
and three two-loop contributions $\tilde T_i(x_1,x_2)$. These
diagrams are given in appendix D. Here we will focus on the
contributions to linear order in $\lambda$.

It is convenient to define the Fourier transform of the diagrams,
which is relevant when computing the corrections to the power
spectrum
 \beq
 T^{(i)}(x_1,x_2)=
 \int\frac{d^3k}{(2\pi)^3}e^{i\vec{k}\cdot(\vec{x}_1-\vec{x}_2)}T^{(i)}(\eta_0,k)~.
 \eeq
In this notation the one loop corrected power spectrum becomes
 \beq \label{pspect}
\mathcal{P}(\eta_0,k) = \frac{k^3}{2\pi^2}\sum_i
T^{(i)}(\eta_0,k)~.
 \eeq

Let us evaluate specifically the diagram which is first order in
$\lambda$. That is the \textit{seagull} diagram $T^{(1)}(x_1,x_2)$
shown in fig.(2). In Fourier space we obtain
 \beq
T^{(1)}(\eta_0,k) = -2
\int_{\eta_{infl}}^{\eta_0}\frac{d\eta}{\eta^4H^4}\textrm{Im}\left[G^{>}_k(\eta_0,\eta)G^{>}_k(\eta_0,\eta)\right]\times
45\lambda\int\frac{d^3k'}{(2\pi)^3}G^{>}_{k'}(\eta,\eta)
 \eeq

\begin{figure}[!hbtp] \label{fig2}
\begin{center}
\includegraphics[width=5.3cm]{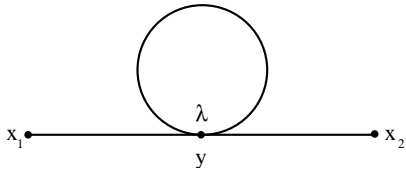}
\end{center}
\caption{The seagull diagram contributing to first order in
$\lambda$.}
\end{figure}

To compute the seagull diagram $T^{(1)}(\eta_0,k)$ given by
 \beq \label{T1}
T^{(1)}_2(\eta_0,k) = -90\lambda
\int_{\eta_{infl}}^{\eta_0}\frac{d\eta}{\eta^4H^4}\textrm{Im}\left[U_k^2(\eta)U_k^{*2}(\eta')\right]\int\frac{d^3k'}{(2\pi)^3}G^{>}_{k'}(\eta,\eta)~,
 \eeq
we have to understand the behavior of the integral. We will first
make some approximations, such that we can solve it analytically.
Since the physically relevant (observable) $k$-modes only spent
less than $60$ e-folds outside the horizon, as a first
approximation, we will take
 \beq \label{U32}
U_k(\eta) = \frac{iH}{k\sqrt{2k}}(1+ik\eta)e^{-ik\eta}~.
 \eeq
In appendix C we have shown the computation, when the mode
function is taken to be given by a Hankel function with index
$\nu\neq 3/2$ and reproduced the correct scaling for the seagull
contribution. We have shown that we are making a very small error
above, when estimating the magnitude of the seagull contribution,
by approximating the mode function with the scale invariant one.

Changing variables to $x=-k\eta$ and letting $\eta_{infl}\to
-\infty$, we obtain the following integral
 \bea \label{T11}
T^{(1)}_1(\eta_0,k) &=&-45
\frac{\lambda}{k^3}\frac{H^2}{4\pi^2}\Delta_N\int_{x_0}^{\infty}\frac{dx}{x^4}\left[(-1+x^2+x_0^2-x_0^2x^2+4x_0x)\sin(2(x-x_0))\right.\nonumber\\
& &\left.+2(-x_0+x_0x^2+x-x_0^2x)\cos(2(x-x_0))\right]~.
 \eea
Since the integrand falls of as a power on sub-Hubble scales
$x>>1$, the dominant contribution to the integral is from
$x\lesssim 1$ and since
 \beq
\int\frac{d^3k'}{(2\pi)^3}G^{>}_{k'}(\eta,\eta)=\frac{H^2}{2\pi^2}\Delta_N~,
 \eeq
is almost constant on super-Hubble scales for physically
observable $k$-modes, we have approximated it with a constant in
the integral in eq.(\ref{T11}). Now the integral is easy to solve
and it yields
 \bea
T^{(1)}_1(\eta_0,k) &=&
-45\frac{\lambda}{k^3}\frac{H^2}{4\pi^2}\Delta_N\left\{\frac{20}{3}\nonumber\right.\\
& &+\left[\frac{1}{3}x_0^2\pi-\frac{1}{3}\pi
+\frac{2}{3}\textrm{Si}(2x_0)-\frac{2}{3}x_0^2\textrm{Si}(2x_0)-\frac{44}{3}x_0\textrm{Ci}(2x_0)\right]\sin(2x_0)\nonumber\\
&
&\left.+\left[-\frac{22}{3}x_0\pi+\frac{44}{3}x_0\textrm{Si}(2x_0)-\frac{2}{3}x_0^2\textrm{Ci}(2x_0)+\frac{2}{3}\textrm{Ci}(2x_0)\right]
\cos(2x_0)\right\}~.
 \eea
Taking the limit $x_0\to 0$, one finds
 \beq
T^{(1)}(\eta_0,k)\simeq -45
\frac{\lambda}{k^3}\frac{H^2}{4\pi^2}\Delta_N\left[\frac{20}{3}+\frac{2}{3}\textrm{Ci}(2x_0)\right]~,
 \eeq
which shows that this one loop contribution to first order in
$\lambda$ is constant on super-Hubble scales up to a logarithmic
time-dependence, as we would have expected within this
approximation.

From eq.(\ref{pspect}) we find on super-Hubble scales
 \beq
\mathcal{P}(\eta_0,k) \simeq
\frac{H^2}{4\pi^2}\left[1-\frac{15}{\pi^2}\lambda\Delta_N\left(10+\textrm{Ci}(-2k\eta_0)\right)+\mathcal{O}(\lambda^2)\right]~.
 \eeq

To estimate the logarithmic correction to the power spectrum, we
can again use $H\simeq 10^{-5}M_p$, $\lambda\simeq 10^{-14}$,
$H_i^4\lesssim 10^{-9}M_p^4$, which implies
 \beq
\frac{45}{3\pi^2}\lambda\Delta_N =
\frac{15\sqrt{3}}{16\pi^2}\sqrt{\lambda}\frac{H_i^4}{M_pH^3}\lesssim
0.016~.
 \eeq
Again we find that within the present approximation, the one loop
correction to the power spectrum is maximally of order $1\%$, if
inflation is sufficiently long. We might also note again, that in the
limit $H_i\to H$ the contribution becomes negligible as expected.
We conclude that our results are consistent with a maximal effect
of a few percent.

\section{Discussion}

We have calculated the one loop corrections to a $\lambda\phi^4$
type of chaotic inflation. We found that in the effective field
theory the quantum corrections are not always suppressed by a
factor $H^2/M_p^2$ as generally expected. In scenarios with very
long inflation there can be a significant enhancement of the one
loop effects. This is because the phase-space volume of $IR$
modes, to be integrated over in the loops, grows with the total
number of e-folds of inflation. Hence, the loops are effectively
suppressed only by a factor of $H^4_i/(H^2M^2_p)$, where $H_i$ is
the Hubble rate at the beginning of inflation. We estimated that
if inflation is very long, or equivalently the Hubble scale at the
beginning of inflation was very close to the Planck scale, the
effects can be as large as perhaps a few percent.

Since the effect comes from an integration of $IR$ modes in the
loop integrals, we expect that it looks similar to a redefinition
of the background parameters of the theory up to a small time
dependence. However, there is no reason a priori why a mass term
of the same magnitude as the one loop induced mass term should
present at tree-level. Such a mass term could affect the slow-roll
parameters and the cosmological scalar-to-tensor consistency
relation, and one can speculate about the theoretical possibility
of an observational indirect evidence for very long inflation.

One aspect which we find also deserves further exploration is if
there could be other new effects from an $IR$ enhancement of one
loop corrections. For instance if it could lead to an enhancement
of the effects found in
\cite{transpl,noncomm,newphys,minlength,eff,collins}. It is as
well tempting to speculate about non-gaussianities, which will
also be enhanced by one loop effects in models with long
inflation. In an effective $\lambda\phi^4$ theory, we have to go
to second order in $\lambda$ to have a one loop integral that
contributes to the three-point function, so it will be suppressed
with a factor of $\lambda$ compared to the two-point seagull
contribution. Going to a $\lambda\phi^6$ type interaction, the
suppression is $H^2/M_p^2$ compared to the two-point seagull
contribution, from dimensional considerations.

%%%%%%%%%%%%%%%%%%%%%%%%%%%%%%%%%%%%%%%%%%%%%%%%%%%%%%
%%%%%%%%%%%%%%%%%%%%%%%%%%%%%%%%%%%%%%%%%%%%%%%%%%%%%%
\overskrift{Acknowledgments}

%%%%%%%%%%%%%%%%%%%%%%%%%%%%%%%%%%%%%%%%%%%%%%%%%%%%%%
%%%%%%%%%%%%%%%%%%%%%%%%%%%%%%%%%%%%%%%%%%%%%%%%%%%%%%
\noindent I would like to thank Kari Enqvist, Massimo Giovannini,
Steen Hannestad, Nemanja Kaloper, Massimo Porrati and David Seery for
discussions and comments. Especially I would like to thank Nemanja
Kaloper for suggesting to explore loop effects in long inflation.
The work was supported in part by the DOE Grant DE-FG03-91ER40674.

%%%%%%%%%%%%%%%%%%%%%%%%%%%%%%%%%%%%%%%%%%%%%%%%%%%%%%
%%%%%%%%%%%%%%%%%%%%%%%%%%%%%%%%%%%%%%%%%%%%%%%%%%%%%%

\appendix

\section{Inflaton perturbations in the ADM formalism}

It is convenient to use the ADM formalism \cite{Arnowitt:1962hi}
to derive the action for the inflaton perturbations. Let us
consider the scalar action of the inflaton field
 \beq
S= \frac{1}{2}\int\sqrt{g}\left[R-(\p\phi)^2-2V(\phi)\right]~,
 \eeq
in the ADM metric, given by
 \beq
ds^2 =
-\mathcal{N}^2dt^2+h_{ij}(dx^i+\mathcal{N}^idt)(dx^j+\mathcal{N}^jdt)~.
 \eeq
In this metric, the action becomes \cite{Arnowitt:1962hi}
 \beq
S=\frac{1}{2}\int\sqrt{h}\left[\mathcal{N}R^{(3)}-2\mathcal{N}V+\mathcal{N}^{-1}\left(E_{ij}E^{ij}-E^2\right)+\mathcal{N}^{-1}
\left(\dot\phi-\mathcal{N}^i\p_i\phi\right)^2-\mathcal{N}h^{ij}\p_i\phi\p_j\phi\right]~,
 \eeq
where
 \beq
E_{ij}=\frac{1}{2}\left(\dot h_{ij}-\nabla_i
\mathcal{N}_j-\nabla_j \mathcal{N}_i\right)~.
 \eeq
As mentioned in the introduction, we find it convenient to discuss
the effective action of the inflaton perturbations in the uniform
curvature gauge, where, when ignoring vector and tensor modes, we have
 \beq
\phi = \phi_c+\delta\phi~,\qquad h_{ij}=a^2\delta_{ij}~,\qquad
\mathcal{N}=1+\al~,\qquad \mathcal{N}^i=\p_i\chi~.
 \eeq
The strength of the ADM formalism, is that the constraint
equations are easily obtained by varying the action in $N$ and
$N_i$, which acts as Lagrange multipliers. In this way the
constraint equations in the uniform curvature gauge becomes
 \beq
-a^2\delta^{ij}\p_i\phi\p_j\phi-2V-\mathcal{N}^{-2}\left(E_{ij}E^{ij}-E^2+\left(\dot\phi-\mathcal{N}^i\p_i\phi\right)^2\right)=0~,
 \eeq
and
 \beq
\nabla_j\left[\mathcal{N}^{-1}\left(E^i_j-\delta^i_jE\right)\right]
=
\mathcal{N}^{-1}\left(\dot\phi-\mathcal{N}^j\p_j\phi\right)\p_i\phi~.
 \eeq
If we perturb the action by taking
 \beq
\phi=\phi_c+\delta\phi~,\qquad \al=\al_1+\al_2+\dots~,\qquad
\chi=\chi_1+\chi_2+\dots~,
 \eeq
and solve the constraints equations order by order, one finds to
first order \cite{Maldacena:2002vr}
 \beq \label{a1x1}
\al_1 = \frac{1}{2}\frac{\dot\phi_c}{H}\delta\phi~,\qquad
\p^2\chi_1=-\frac{1}{2}\frac{\dot\phi_c}{H}\dot{\delta\phi}-\frac{1}{2}\dot\phi_c\frac{\dot
H}{H}\delta\phi+\frac{1}{2}\frac{\ddot\phi}{H}\delta\phi~.
 \eeq
Generally, in order to obtain the action to order $n$, we need
only to derive the constraint equations to order $n-1$, since the
$n$'th order terms multiplies the constraint equation to zero'th
order. In fact, it also turns out that to obtain the action to
third order in perturbations, one only needs the first order terms
in eq.(\ref{a1x1}), since $\al_2$, $\chi_2$ cancels out to leading
order in the slow-roll expansion. One obtains
 \bea \label{S3}
S_3 &=& \int
a^3\left[-\frac{1}{4}\frac{\dot\phi_c}{H}\dot{\delta\phi}^2\delta\phi-\frac{a^2}{4}\frac{\dot\phi_c}{H}\delta\phi(\p\delta\phi)^2-\dot{\delta\phi}\p_i\chi_1\p_i\delta\phi
\right.\nonumber\\
& & +\frac{3}{8}\frac{\dot\phi_c^3}{H}\delta\phi^3
-\frac{1}{4}\frac{\dot\phi_c}{H}V''\delta\phi^3-\frac{1}{6}V'''\delta\phi^3+\frac{1}{4}\frac{\dot\phi_c^3}{H^2}\delta\phi^2\dot{\delta\phi}
+\frac{1}{4}\frac{\dot\phi_c^2}{H}\delta\phi^2\p^2\chi_1\nonumber\\
&
&~\left.+\frac{1}{4}\frac{\dot\phi_c}{H}\left(-\delta\phi\p_i\p_j\chi_1\p_i\p_j\chi_1+\delta\phi\p^2\chi_1\p^2\chi_1\right)\right]~,
 \eea
as first derived by Maldacena \cite{Maldacena:2002vr}, and
subsequently generalized in
Ref.~\cite{Creminelli:2003iq,Seery:2005wm,Seery:2005gb}. By going
one order further, we can in a similar fashion obtain the action
to fourth order in perturbations
 \bea \label{S4}
S_4 &=& \int
a^3\left[\frac{1}{16}\frac{\dot\phi_c^2}{H^2}\dot{\delta\phi}^2\delta\phi^2-\frac{a^2}{16}\frac{\dot\phi_c^2}{H^2}\delta\phi^2(\p\delta\phi)^2+\frac{1}{2}\frac{\dot\phi_c}{H}\delta\phi\dot{\delta\phi}\p_i\chi_1\p_i\delta\phi+\frac{1}{24}\frac{\dot\phi_c^3}{H^2}\delta\phi^3\p^2\chi_1
\right.\nonumber\\
& &-\frac{15}{64}\frac{\dot\phi_c^4}{H^2}\delta\phi^4
-\frac{1}{16}\frac{\dot\phi_c^2}{H^2}V''\delta\phi^4-\frac{1}{12}\frac{\dot\phi_c}{H}V'''\delta\phi^4-\frac{1}{24}V''''\delta\phi^4+\frac{1}{2}(\p_i\chi_1)^2(\p_j\delta\phi)^2\nonumber\\
 & &~\left.-\dot{\delta\phi}\p_i\chi_2\p_i\delta\phi-\frac{1}{16}\frac{\dot\phi_c^2}{H^2}\left((\p^2\chi_1)^2-(\p_i\p_j\chi_1)^2\right)\delta\phi^2-\frac{1}{2}\left((\p^2\chi_2)^2-(\p_i\p_j\chi_2)^2\right)
\right.\nonumber\\
&
&\left.
+\frac{\dot\phi_c}{2H}\left(\delta\phi\p^2\chi_1\p^2\chi_2-\delta\phi\p_i\p_j\chi_1\p_i\p_j\chi_2\right)-\p^2\chi_1\p^2\chi_3+\p_i\p_j\chi_1\p_i\p_j\chi_3\right.\nonumber\\
& & \left.
-\left(2H\p^2\chi_2+\frac{1}{2}V''\delta\phi^2+\frac{1}{2}(\p\delta\phi)^2+\frac{1}{4}\frac{\dot\phi_c^2}{H^2}V\delta\phi^2+VF(\delta\phi,\dot\delta\phi)\right.\right.\nonumber\\
 & &\left.\left.
-\frac{1}{2}(\p^2\chi_1)^2+\frac{1}{2}(\p_i\p_j\chi_1)^2
+\frac{1}{2}\dot{\delta\phi}^2-\dot\phi_c\p_i\chi_1\p_i\delta\phi\right)F(\delta\phi,\dot\delta\phi)\right]~,
 \eea
valid up to total derivative terms. We also note that to leading
order in slow-roll $\al_3$ has cancelled out of the action. Above,
we also used the solution to the constraint equations to second
order,
 \beq
\al_2=\frac{\dot\phi_c^2}{8H^2}+F(\delta\phi,\dot{\delta\phi})~,
 \eeq
and
 \bea
\p^2\chi_2 &=&
\frac{3}{8}\frac{\dot\phi_c^2}{H}\delta\phi^2+\frac{3}{4}\frac{\ddot\phi_c}{H\dot\phi_c}\delta\phi^2-\frac{a^2}{4H}(\p\delta\phi)^2-\frac{1}{4H}\dot{\delta\phi}^2+\frac{\dot\phi_c}{2H}\p_i\chi_1\p_i\delta\phi\nonumber\\
& &+\frac{1}{4H}\left((\p^2\chi_1)^2-(\p_i\p_j\chi_1)^2\right)-\frac{V}{H}F(\delta\phi,\dot{\delta\phi})~,
 \eea
where we have for convenience defined
 \beq
F(\delta\phi,\dot{\delta\phi})=
\frac{1}{2H}\p^{-2}\left[\p^2\al_1\p^2\chi_1-\p_i\p_j\al_1\p_i\p_j\chi_1+\p_i\dot{\delta\phi}
\p_i\delta\phi+\dot{\delta\phi}\p^2\delta\phi\right]~.
 \eeq

The extra terms in the $S_4$ action involving $\al_3$, which cancels to leading order in slow-roll, are $
(2\al_1\al_2-\al_3)(4H\p^2\chi_1+2\dot\phi_c\dot{\delta\phi})$, $-(V'+\frac{\dot\phi_c}{H}V)\delta\phi\al_3~.$ By using $V$ to leading order in slow-roll we also neglected a term $(5/128)(\dot\phi_c^6/H^4)\delta\phi^4$ in the action $S_4$, and in a similar fashion we have also neglected terms of higer order in slow-roll in the expression for $\chi_2$.

\section{IR scaling behavior}

In this section of the appendix, we will evaluate the $IR$ part of
the correlator $\lmk \delta\phi^2\rmk_0$. We know that in the
chosen gauge $|U_k| = (\dot\phi/H)\mathcal{R}_k$, where
$\mathcal{R}_k= -\zeta_k$ is the gauge invariant curvature
perturbation, constant on super-Hubble scales. Thus, we can solve
the mode equations in pure de Sitter ($\ep=0$) and then evaluate
$\mathcal{R}_k$, but using the value of $H=H_k$ and
$\dot\phi=\dot\phi_k$ when the given mode crosses the horizon.
Since we know that $\mathcal{R}_k$ and $U_k$ scales in the same
way, we can first integrate over the power spectrum of
$\mathcal{R}_k$ and then relate that to the two-point function of
$\delta\phi$ at the end. In pure de Sitter on super-Hubble scales
$|U_k(\eta)|=H/2\pi$, so the curvature perturbation in this
approach can be written as
 \beq
\mathcal{R}_k\simeq
\frac{1}{2\pi}\left(\frac{H_k^2}{\dot\phi_k}\right)~.
 \eeq
If we specify to the specific $\lambda\phi^4$ model of chaotic
inflation, this implies that we obtain the right $k$ scaling from
 \beq
U_k(\eta) = \lmk\frac{H_k}{H(\eta)} \rmk^{3/2}U^{ds}_k(\eta)=
\frac{\sqrt{\pi}}{2}H\eta^{3/2}\lmk\frac{H_k}{H(\eta)}
\rmk^{3/2}H_{3/2}^{(2)}(k\eta)~,
 \eeq
where the mode solution in pure de Sitter was denoted by
$U^{ds}_k(\eta)$.

If we let $N=\ln(a/a_i)$ denote the number of e-foldings of
expansions since the beginning of inflation, it is convenient to
recast the relevant integral into an integral over $N$. To do so,
we use the relations $d\ln k= dN$, $\ln(aH/k)=N$ and
 \beq
H_k^2 =\frac{\lambda}{12M_p^2}\left(\phi_i^2-64M_p^2N\right)^2~.
 \eeq
Now we can easily evaluate the dominant part of the $IR$
contribution to the relevant correlator
 \bea
\lle\right. \lb \delta\phi(t,{\bf x}) \rb^2\left.\rgr_{IR} &\simeq
& \frac{H^2}{4\pi^2}\int_{a_iH_i}^{aH}
\frac{dk}{k}\lmk\frac{H_k}{H} \rmk^{3} \nonumber\\ &=&
\frac{H^2}{4\pi^2}\lmk\frac{\lambda}{12M_p^2H^2}\rmk^{3/2}\int_0^N
dN \left(\phi_i^2-64M_p^2N\right)^3
 \eea
Using $N\simeq \phi_i^2/(64M_p^2)$, we obtain
 \bea
\lle\right. \lb \delta\phi(t,{\bf x}) \rb^2\left.\rgr_{IR}
&\simeq&
\frac{H^2}{4\pi^2}\lmk\frac{\lambda}{12M_p^2H^2}\rmk^{3/2}\frac{1}{256
M_p^2}\phi_i^8\nonumber\\
&=&\frac{H^2}{16\pi^2}\lmk\lambda\frac{1024}{3}\frac{M_p^2}{H^2}\rmk^{3/2}
N^4~. \label{corr2}
 \eea
As mentioned also in section (2.1), our results are consistent
with \cite{Mukhanov:1996ak,Abramo:1997hu,Afshordi:2000nr}, when
comparison is possible.

\section{Seagull integral}

Consider the seagull integral in eq.(\ref{T1}). Instead of using
the approximation in eq.(\ref{U32}), we want to use the exact
expression with the correct scaling behavior
 \beq
U_k(\eta) =
\frac{\sqrt{\pi}}{2}H(-\eta)^{3/2}H_{\nu}^{(2)}(-k\eta)~.
 \eeq
In the approximation where we can treat $H^2\Delta_N$ as constant,
the relevant integral becomes
 \bea
I_1(\eta_0,k) &=&
\int\frac{d\eta}{\eta^4H^4}\textrm{Im} \left[U_k^2(\eta_0)U_k^{*2}(\eta)\right]\nonumber\\
&=& \frac{\pi^2}{16}\int d\eta ~\eta^2
\textrm{Im}\left[H_{\nu}^{(2)}(-k\eta_0)H_{\nu}^{(2)}(-k\eta_0)H_{\nu}^{(1)}(-k\eta)H_{\nu}^{(1)}(-k\eta)\right]~.
 \eea
It is useful to make a transformation to a dimensionless parameter
$x=-k\eta$, such that the integral becomes
 \beq
I_1(x_0,k) = -\frac{\pi^2}{16}\int dx ~x^2
\textrm{Im}\left[H_{\nu}^{(2)}(x_0)H_{\nu}^{(2)}(x_0)H_{\nu}^{(1)}(x)H_{\nu}^{(1)}(x)\right]~.
 \eeq
We know from our exact calculation in the case $\nu =3/2$, that in
the limit $x_0\to 0$, the dominant contribution to the integral
are a constant contribution from $x\simeq 1$ and a logarithmic
$x_0$ dependent contribution from $x<<1$. That means, that we can
obtain a good approximation to the integral by expanding the
Hankel function in the small argument limit, reproducing the $x_0$
dependent contribution to the integral, and add the constant
constant contribution computed in the $\nu=3/2$ case. In this way
we obtain
 \bea
I_1(x_0,k) &=&
\frac{1}{k^3}\frac{\pi^2}{8}\frac{2^{2\nu}}{\sin^3(\nu\pi)\Gamma^3(1-\nu)\Gamma(\nu+1)}\left(\frac{1}{3}
-\frac{1}{3-2\nu}\right)x_0^{3-2\nu}+\frac{20}{12}~,\qquad
\textrm{for}\qquad
\nu\neq 3/2~,\nonumber\\
I_1(x_0,k) &=&
\frac{1}{k^3}\frac{-\pi^2}{8}\frac{2^{2\nu}}{\Gamma^3(1-\nu)\Gamma(\nu+1)}\left(\frac{1}{3}
-\ln(x_0)\right)+\frac{20}{12}~,\qquad \textrm{for}\qquad \nu
=3/2~. \label{I1appr}
 \eea
In fig.(4), we have compared the exact solution for the integral
in the case $\nu=3/2$ with the approximate solution given in
eq.(\ref{I1appr}). We find that the approximation is reasonably
good on very super-Hubble scales $x_0<<10^{-3}$.

\begin{figure}[!hbtp] \label{fig3}
\begin{center}
\includegraphics[width=6.3cm]{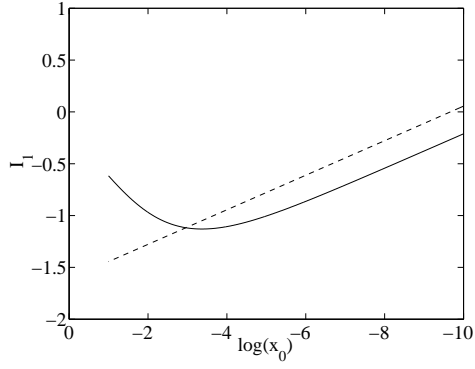}
\end{center}
\caption{Different approximations to the integral $I_1(x_0)$ in
$T^{(1)}(k)$. The solid line is the solution calculated in section
3.2 and the dashed line is the solution calculated above in
eq.(\ref{I1appr}).}
\end{figure}

\section{One loop diagrams}

The Bubble diagram, $T^{(2)}_1$ shown in fig.(4), is given by
 \bea
T^{(2)}_1(k,\eta_0) &=&
36\lambda^2\int_{-\infty}^{\eta_0}d\eta\int\frac{d^3y}{\eta^4H^4}\int_{-\infty}^{\eta_0}d\eta'\int\frac{d^3y'}{\eta'^4H'^4}\phi_c(\eta)\phi_c(\eta')
\left[\left(G^>(x_1,y)-G^<(x_1,y)\right)\right. \nonumber\\
&\times
&\left.\left(G^>(x_2,y')G^>(y,y')G^>(y,y')-G^<(x_2,y')G^<(y,y')G^<(y,y')\right)\right]~.
 \eea
The diagram $T^{(2)}_2$, shown in fig.(5), is
 given by
 \bea
T^{(2)}_2(k,\eta_0) &=&
24(48\lambda)^2\int_{-\infty}^{\eta_0}d\eta\int\frac{d^3y}{\eta^4H^4}\int_{-\infty}^{\eta_0}d\eta'\int\frac{d^3y'}{\eta'^4H'^4}\phi_c(\eta)\phi_c(\eta')\nonumber\\
&\times&
\left[\left(G^>(x_1,y)G^>(x_2,y')-G^<(x_1,y)G^<(x_2,y')\right)\right. \nonumber\\
&\times
&\left.\left(G^>(y,y')G^>(y',y')-G^<(y,y')G^<(y',y')\right)\right]~.
 \eea
However, since the tadpole subgraph is canceled by the tadpole
renormalization condition, this diagram will also cancel out.

Finally there is three two loop diagrams contributing to second
order in $\lambda$. They are labeled  $\tilde T^{(2)}_1$, $\tilde
T^{(2)}_2$, $\tilde T^{(2)}_3$ in fig.(7), and are given below
 \bea
\tilde T^{(2)}_1(k,\eta_0) &=&
3(48\lambda)^2\int_{-\infty}^{\eta_0}d\eta\int\frac{d^3y}{\eta^4H^4}\int_{-\infty}^{\eta_0}d\eta'\int\frac{d^3y'}{\eta'^4H'^4}\nonumber\\
&\times&
\left[\left(G^>(x_1,y)G^>(x_2,y')-G^<(x_1,y)G^<(x_2,y')\right)\right. \nonumber\\
&\times
&\left.\left(G^>(y,y')G^>(y,y')G^>(y',y')-G^<(y,y')G^<(y,y')G^<(y',y')\right)\right]~.
 \eea
 \bea
\tilde T^{(2)}_2(k,\eta_0) &=&
4(48\lambda)^2\int_{-\infty}^{\eta_0}d\eta\int\frac{d^3y}{\eta^4H^4}\int_{-\infty}^{\eta_0}d\eta'\int\frac{d^3y'}{\eta'^4H'^4}\nonumber\\
&\times&
\left[\left(G^>(x_1,y)G^>(y,y)-G^<(x_1,y)G^<(y,y)\right)\right. \nonumber\\
&\times
&\left.\left(G^>(x_2,y')G^>(y,y')G^>(y',y')-G^<(x_2,y')G^<(y,y')G^<(y',y')\right)\right]~.
 \eea
 \bea
\tilde T^{(2)}_3(k,\eta_0) &=&
4(48\lambda)^2\int_{-\infty}^{\eta_0}d\eta\int\frac{d^3y}{\eta^4H^4}\int_{-\infty}^{\eta_0}d\eta'\int\frac{d^3y'}{\eta'^4H'^4}\nonumber\\
&\times&
\left[\left(G^>(x_1,y)G^>(y,y')-G^<(x_1,y)G^<(y,y')\right)\right. \nonumber\\
&\times
&\left.\left(G^>(x_2,y')G^>(y,y')G^>(y,y')-G^<(x_2,y')G^<(y,y')G^<(y,y')\right)\right]~.
 \eea
\begin{figure}[!hbtp] \label{fig4}
\begin{center}
\includegraphics[width=5.3cm]{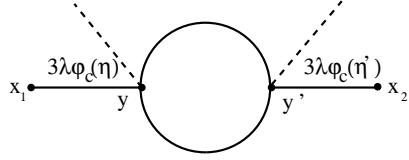}
\end{center}
\caption{The bubble diagram $T^{(2)}_1$.}
\end{figure}
\begin{figure}[!hbtp] \label{fig5}
\begin{center}
\includegraphics[width=5.3cm]{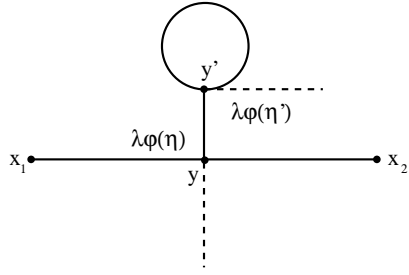}
\end{center}
\caption{The tadpole sub-diagram in the $T^{(2)}$ contribution
above, is canceled to higher order by the tadpole renormalization
condition.}
\end{figure}
\begin{figure}[!hbtp] \label{fig7}
\begin{center}
\includegraphics[width=5.3cm]{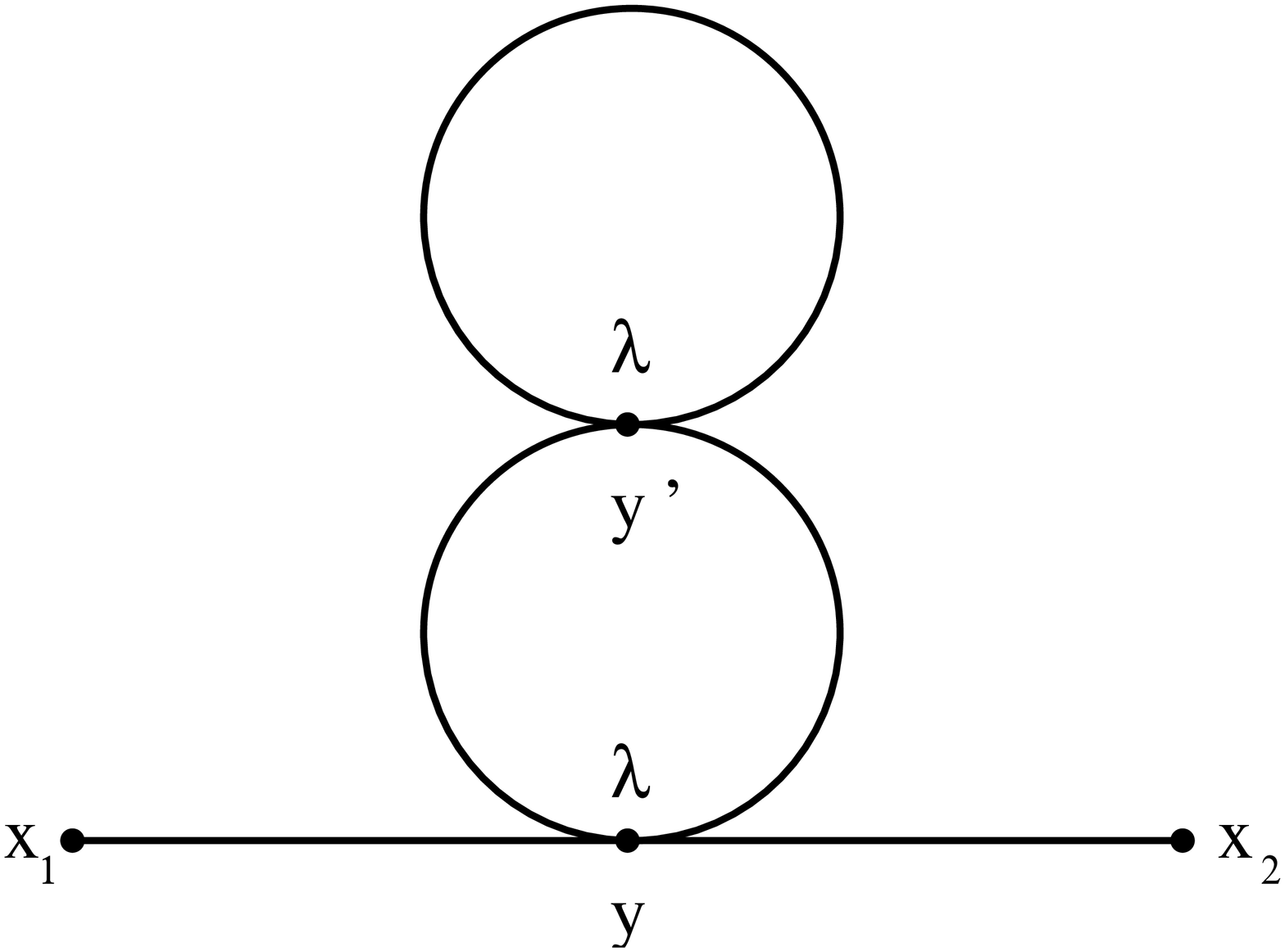}
\includegraphics[width=5.3cm]{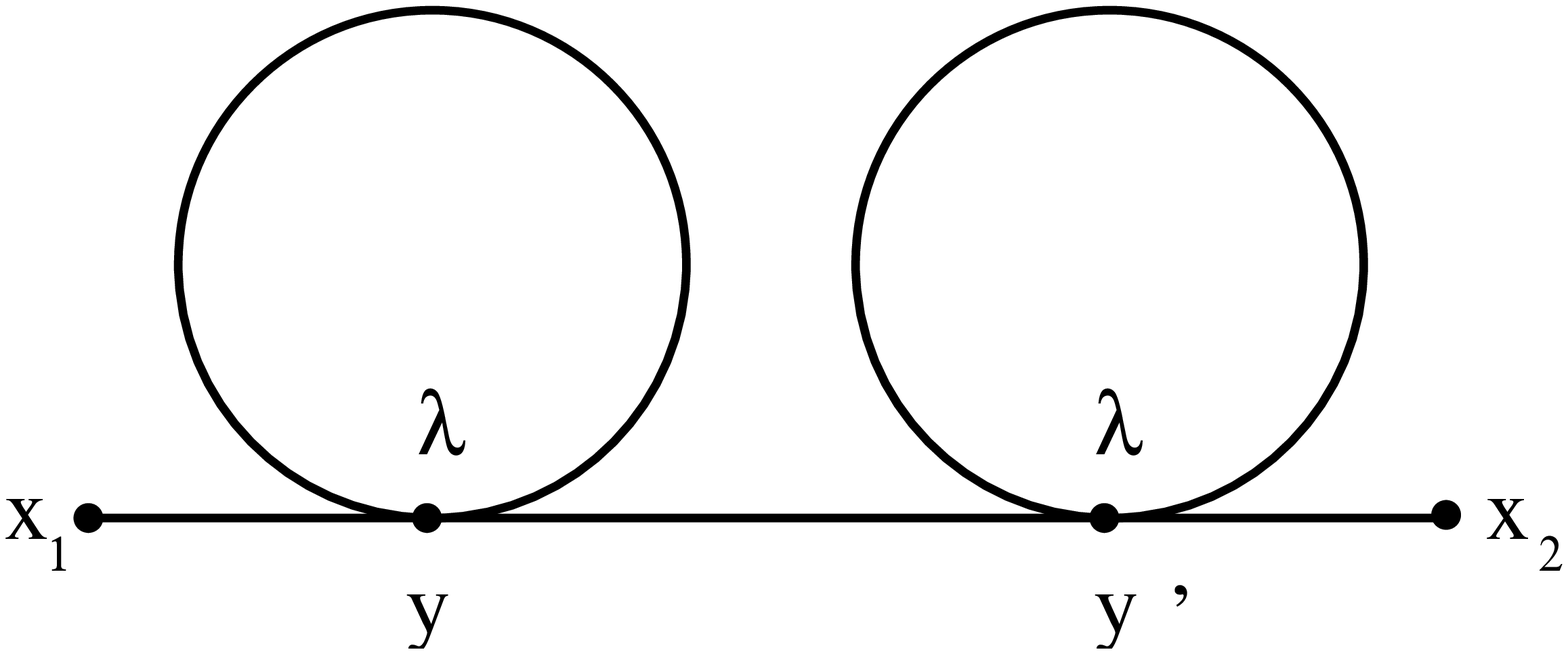}
\includegraphics[width=5.3cm]{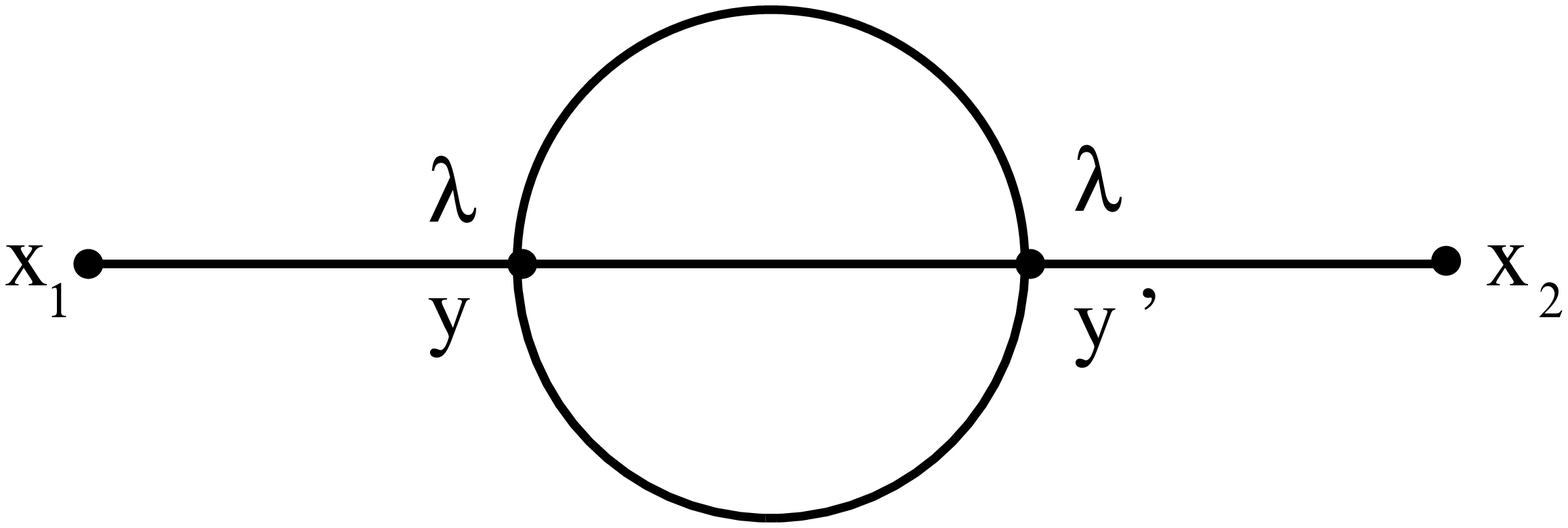}
\end{center}
\caption{The two loop contributions, contributing to second order
in $\lambda$.}
\end{figure}

%\newpage

\end{document}